\documentclass[%
 reprint,
 amsmath,amssymb,
 aps,
pre,
]{revtex4-1}

\usepackage{graphicx}
\usepackage{dcolumn}
\usepackage{bm}
\usepackage{multirow}
\usepackage{longtable}

\newcolumntype{R}[1]{>{\raggedleft\arraybackslash}p{#1}}

\begin{document}


\title{Analysis of food pairing in regional cuisines of India}

\author{Anupam Jain}
 \affiliation{%
 Center for System Science, Indian Institute of Technology Jodhpur, Jodhpur, 342011, India
}%
\author{Rakhi N K}
 \affiliation{%
 Center for Biologically Inspired System Science, Indian Institute of Technology Jodhpur, Jodhpur, 342011, India
}%
\author{Ganesh Bagler}%
\email{bagler@iitj.ac.in; ganesh.bagler@gmail.com}
 \affiliation{%
 Center for Biologically Inspired System Science, Indian Institute of Technology Jodhpur, Jodhpur, 342011, India
}%

\date{\today}

\begin{abstract}
Any national cuisine is a sum total of its variety of regional cuisines, which are the cultural and historical identifiers of their respective regions. India is home to a number of regional cuisines that showcase its culinary diversity. Here, we study recipes from eight different regional cuisines of India spanning various geographies and climates. We investigate the phenomenon of food pairing which examines compatibility of two ingredients in a recipe in terms of their shared flavor compounds. Food pairing was enumerated at the level of cuisine, recipes as well as ingredient pairs by quantifying flavor sharing between pairs of ingredients. Our results indicate that each regional cuisine follows negative food pairing pattern; more the extent of flavor sharing between two ingredients, lesser their co-occurrence in that cuisine. We find that frequency of ingredient usage is central in rendering the characteristic food pairing in each of these cuisines. Spice and dairy emerged as the most significant ingredient classes responsible for the biased pattern of food pairing. Interestingly while individual spices contribute to negative food pairing, dairy products on the other hand tend to deviate food pairing towards positive side. 
Our data analytical study highlighting statistical properties of the regional cuisines, brings out their culinary fingerprints that could be used to design algorithms for generating novel recipes and recipe recommender systems. It forms a basis for exploring possible causal connection between diet and health as well as prospection of therapeutic molecules from food ingredients.
Our study also provides insights as to how big data can change the way we look at food. 

\begin{description}
\item[PACS numbers]
89.75.-k, 82.20.Wt, 87.18.Vf, 87.10.Vg, 89.90.+n
\end{description}
\end{abstract}

\pacs{Valid PACS appear here}
\maketitle


\section{\label{sec:level1}Introduction \\}
Cooking is a unique trait humans possess and is believed to be a major cause of increased brain size~\cite{Navarrete2011,Wrangham2009,Fonseca-Azevedo2012}. While cooking encompasses an array of food processing techniques~\cite{Pollan2014}, cuisine is an organized series of food preparation procedures intended to create tasty and healthy food. India has a unique blend of culturally and climatically diverse regional cuisines. Its culinary history dates back to the early Indus valley civilization~\cite{Weber2011,Kashyap2010,Lawler2012}. Indian dietary practices are deeply rooted in notions of disease prevention and promotion of health.

Food perception involving olfactory and gustatory mechanisms is the primary influence for food preferences in humans. These preferences are also determined by a variety of factors such as culture, climate geography and genetics, leading to emergence of regional cuisines~\cite{Pollan2014,Appadurai2009,Sherman1999,Zhu2013,Birch1999,Ventura2013a}. Food pairing is the idea that ingredients having similar flavor constitution may taste well in a recipe. Chef Blumenthal was the first to propose this idea which in this study we term as positive food pairing~\cite{Blumenthal2008}. Studies by Ahn et al found that North American, Latin American and Southern European recipes follow this food pairing pattern where as certain others like North Korean cuisine and Eastern European cuisines do not~\cite{Ahnert2013,Ahn2011}. Our previous study of food pairing in Indian cuisine revealed a strong negative food pairing pattern in its recipes~\cite{Jain2015}.

Knowing that each of the regional cuisines have their own identity, the question we seek to answer in this paper is whether the negative food pairing pattern in Indian cuisine is a consistent trend observed across all of the regional cuisines or an averaging effect. Towards answering this question, we investigated eight geographically and culturally prominent regional cuisines viz. Bengali, Gujarati, Jain, Maharashtrian, Mughlai, Punjabi, Rajasthani and South Indian. The pattern of food pairing was studied at the level of cuisine, recipes and ingredient pairs. Such a multi-tiered study of these cuisines provided a thorough understanding of its characteristics in terms of ingredient usage pattern. We further identified the features that contribute to food pairing, thereby revealing the role of ingredients and ingredient categories in determining food pairing of the regional cuisines.

Availability of large datasets in the form of cookery blogs and recipe repositories has prompted the use of big data analytical techniques in food science and has led to the emergence of computational gastronomy. This field has made advances through many recent studies~\cite{Ahnert2013,Ahn2011,Varshney2013a,Teng2012} which is changing the overall outlook of culinary science in recent years. Our study is an offshoot of this approach. We use statistical and computational models to analyse food pairing in the regional cuisines. Our study reveals the characteristic signature of each Indian regional cuisines by looking at the recipe and ingredient level statistics of the cuisine.

\section{Results and Discussion}
Details of recipes, ingredients, and their corresponding flavor compounds constitute the primary data required for study of food pairing in a cuisine. Much of this is documented in the form of books and recently through online recipe sources. We obtained the Indian cuisine recipes data from one of the popular cookery websites \emph{TarlaDalal.com}~\cite{Dalal2014}. The flavor profiles of ingredients were compiled using previously published data~\cite{Ahn2011} and through extensive literature survey. Table~\ref{table1} lists details of recipes and ingredients in each of the regional cuisines.

\begin{table}[!ht]
\caption{
{\bf Statistics of regional cuisines}}
\begin{tabular}{|l|l|l|}
\hline
\multicolumn{1}{|l|}{\bf Cuisine} & \multicolumn{1}{l|}{\bf Recipe count} & \multicolumn{1}{|l|}{\bf Ingredient count}\\ \hline
Bengali & 156 & 102 \\ \hline
Gujarati & 392 & 112 \\ \hline
Jain & 447 & 138 \\ \hline
Maharashtrian & 130 & 93 \\ \hline
Mughlai & 179 & 105 \\ \hline
Punjabi & 1013 & 152 \\ \hline
Rajasthani & 126 & 78 \\ \hline
South Indian & 474 & 114 \\ \hline
\end{tabular}
\begin{flushleft} Recipes of size $\geq$ 2 were considered for the purpose of flavor analysis.
\end{flushleft}
\label{table1}
\end{table}

The ingredients belonged to following 15 categories: spice, vegetable, fruit, plant derivative, nut/seed, cereal/crop, dairy, plant, pulse, herb, meat, fish/seafood, beverage, animal product, and flower. Category-wise ingredient statistics of regional cuisines is provided in~\ref{S1_Table}.

\subsection{Statistics of recipe size and ingredient frequency}
\begin{figure}[h]
\centering
\includegraphics[scale=.4]{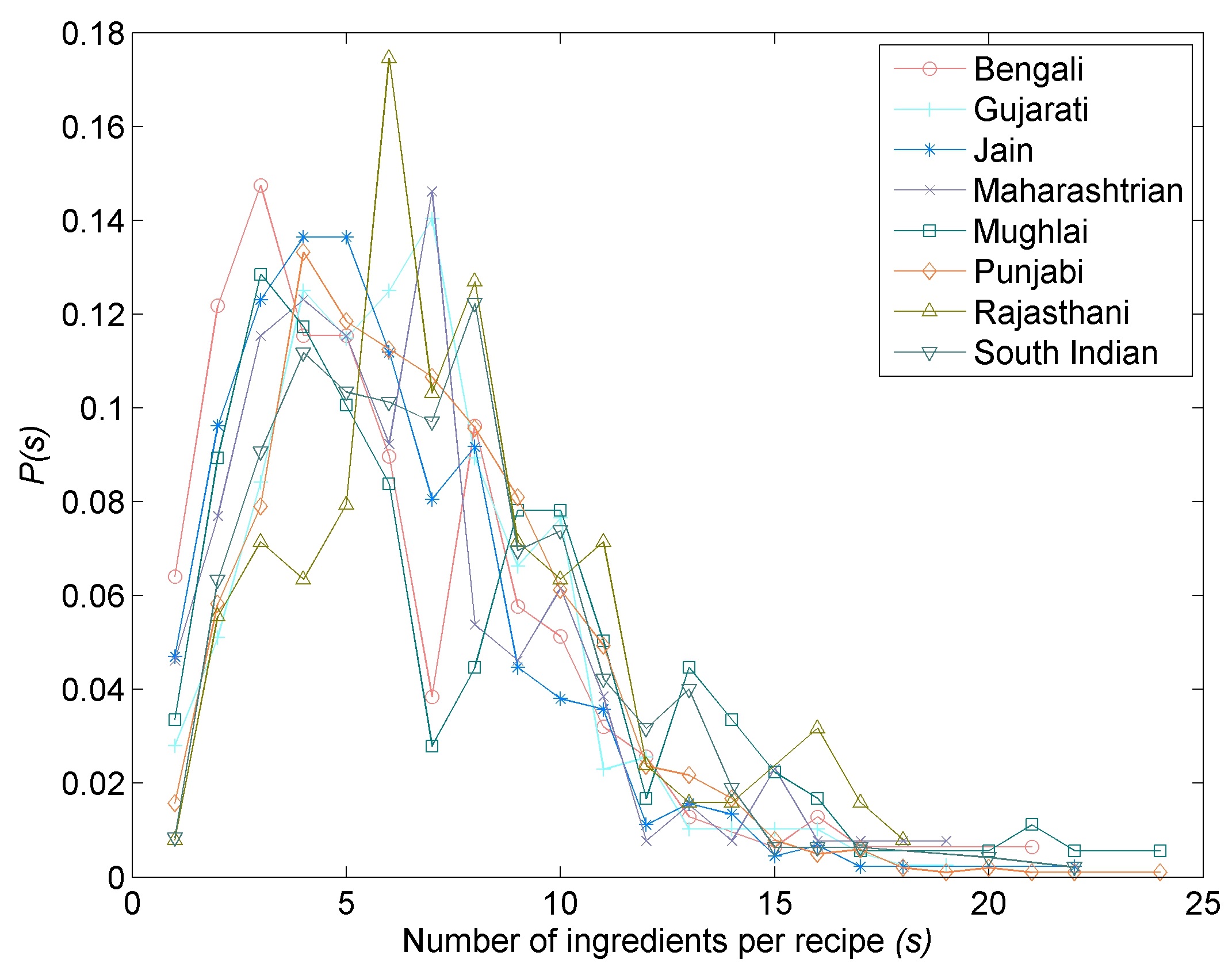}
\caption{{\bf Recipe size distributions.}
Plot of probability of finding a recipe of size $s$ in the cuisine. Consistent with other cuisines, the distributions are bounded. Mughlai and Punjabi cuisines have recipes of large sizes compared to other cuisines.}
\label{Fig_1:size dist}
\end{figure}

\begin{figure}[h]
\includegraphics[scale=.4]{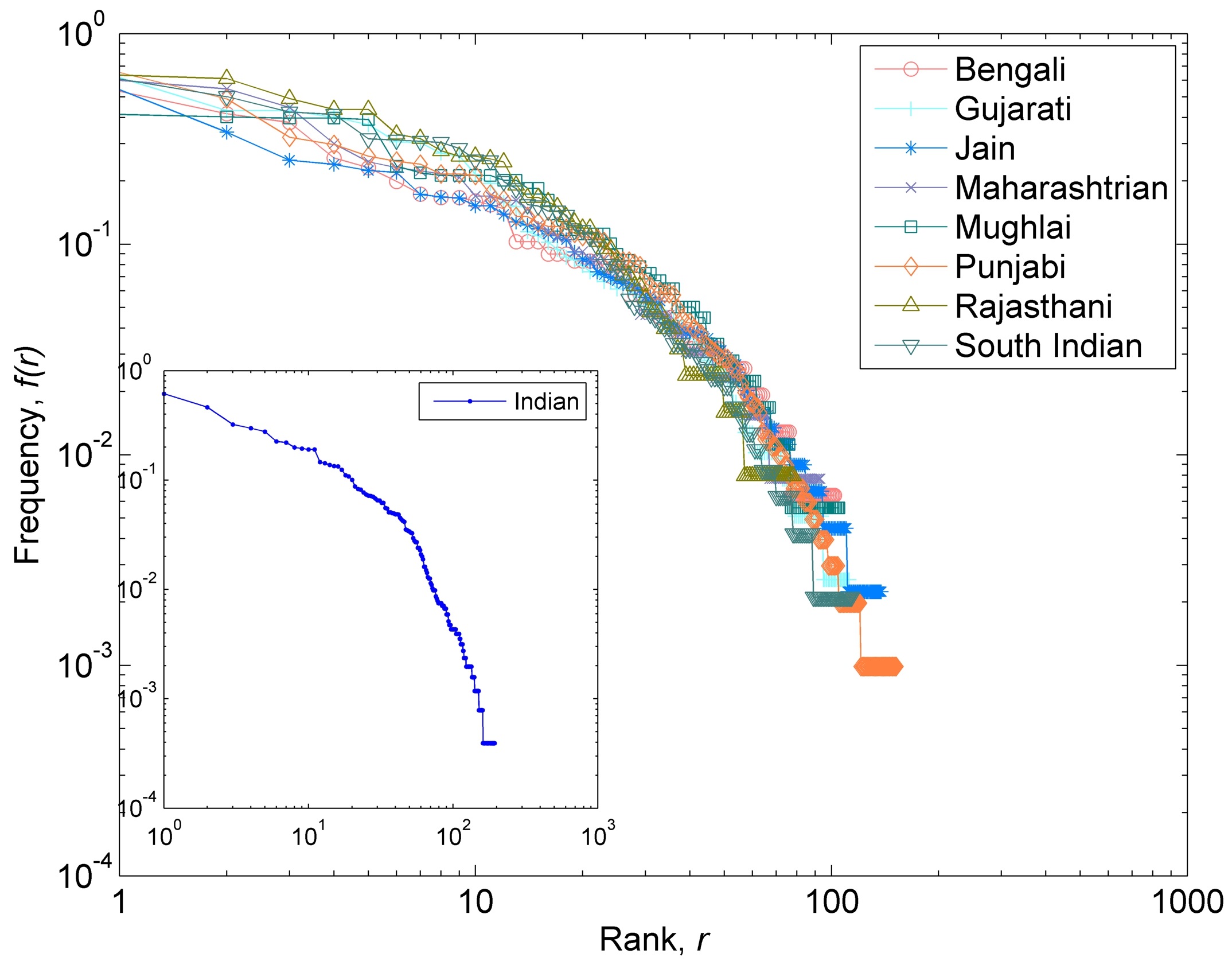}
\caption{{\bf Frequency-Rank distributions.}
Ingredients ranked as per their frequency of use in the cuisine. Higher the occurrence, better the rank of the ingredient. All the cuisines have similar ingredient distribution profile indicating generic culinary growth mechanism. Inset shows the ingredient frequency-rank distribution for the whole Indian cuisine.}
\label{Fig_2:freq-rank}
\end{figure}

We started with investigation of preliminary statistics of regional cuisines. All the eight regional cuisines under consideration showed bounded recipe-size distribution (Figure~\ref{Fig_1:size dist}). 
While most cuisines followed uni-modal distribution, Mughlai cuisine showed a strong bimodal distribution and had recipes with large sizes when compared with the rest. This could be an indication of the fact that Mughlai is derivative of a royal cuisine. To understand the ingredient usage pattern, we ranked ingredients according to decreasing usage frequency within each cuisine. As shown in Figure~\ref{Fig_2:freq-rank}, all cuisines showed strikingly similar ingredient usage profile reflecting the pattern of Indian cuisine (Figure~\ref{Fig_2:freq-rank}, inset). While indicating a generic culinary growth mechanism, the distributions also show that certain ingredients are excessively used in cuisines depicting their inherent `fitness' or popularity within the cuisine.

\subsection{Food pairing hypothesis}
\begin{figure}[h]
\centering
\includegraphics[scale=.3]{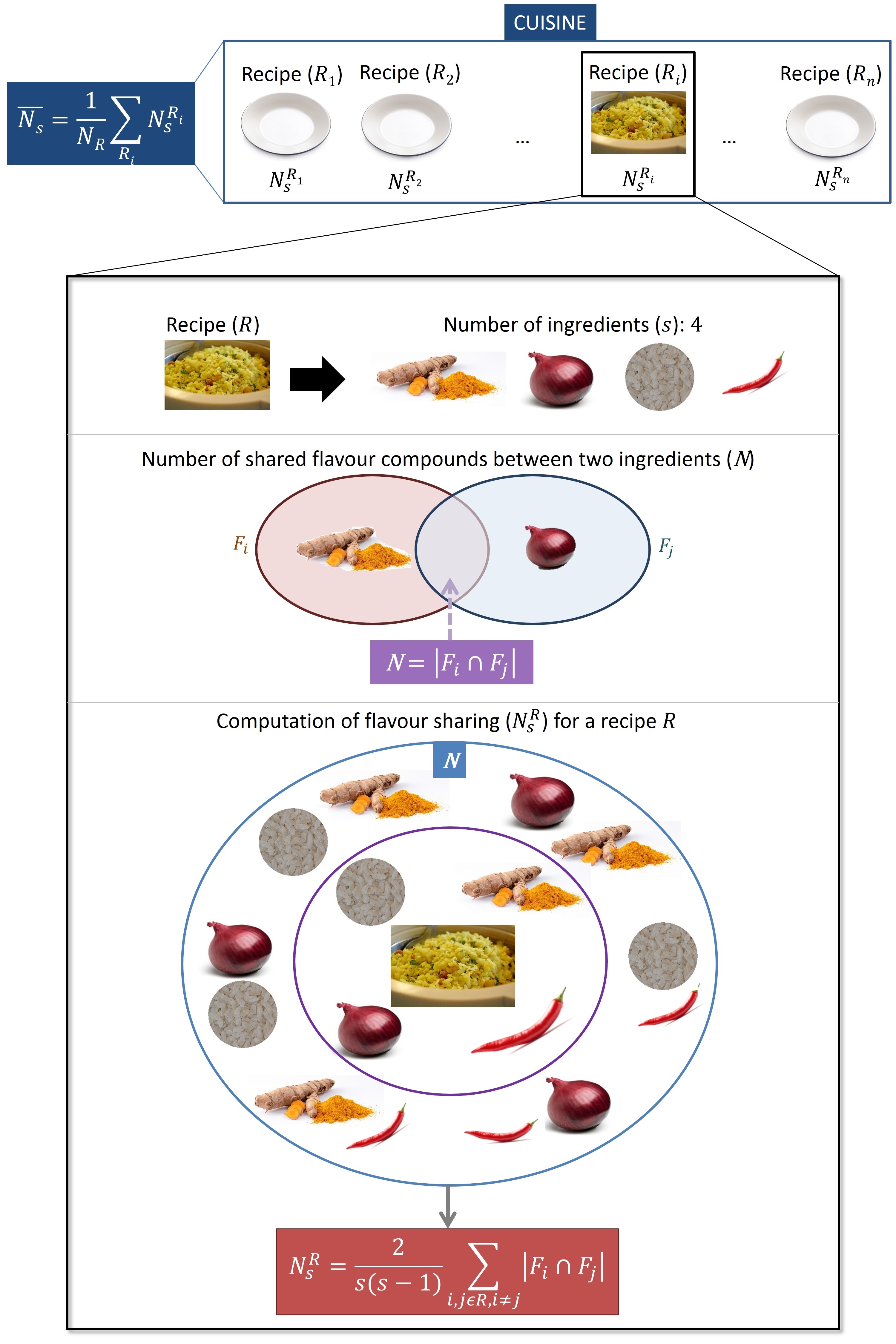}
\caption{{\bf Schematic for calculation of `average $N_s$' ($\overline{N}_s$).} Illustration of procedure for calculating the average $N_s$ for a given cuisine. Beginning with an individual recipe, average $N_s$ of the recipe ($N_s^R$) was calculated. Averaging $N_s^R$ over all the recipes returned $\overline{N}_s$ of the cuisine.
}
\label{Fig_3:Ns calculation}
\end{figure}

Food pairing hypothesis is a popular notion in culinary science. It asserts that two ingredients sharing common flavor compounds taste well when used together in a recipe. This hypothesis has been confirmed for a few cuisines such as North American, Western European and Latin American~\cite{Ahn2011}. In contrast, Korean and Southern European cuisines have been shown to deviate from positive food pairing. Our previous study of food pairing in Indian cuisine at the level of cuisine, sub-cuisines, recipes and ingredient pairs has shown that it is characterized with a strong negative food pairing~\cite{Jain2015}. We quantify food pairing with the help of flavor profiles of ingredients. Flavor profile represents a set of volatile chemical compounds that render the characteristic taste and smell to the ingredient. Starting with the flavor profiles of each of the ingredients, average food pairing of a recipe ($N_s^R$) as well as that of the cuisine ($\overline{N}_s$) was computed as illustrated in Figure~\ref{Fig_3:Ns calculation}. The extent of deviation of $\overline{N}_s$ of the cuisine, when compared to that of a `random cuisine' measures the bias in food pairing. The higher/lower the value of $\overline{N}_s$ from that of its random counterpart the more positive/negative it is. 

\subsection{Regional cuisines of India exhibit negative food pairing}
\begin{figure*}[ht]
\includegraphics[scale=.53]{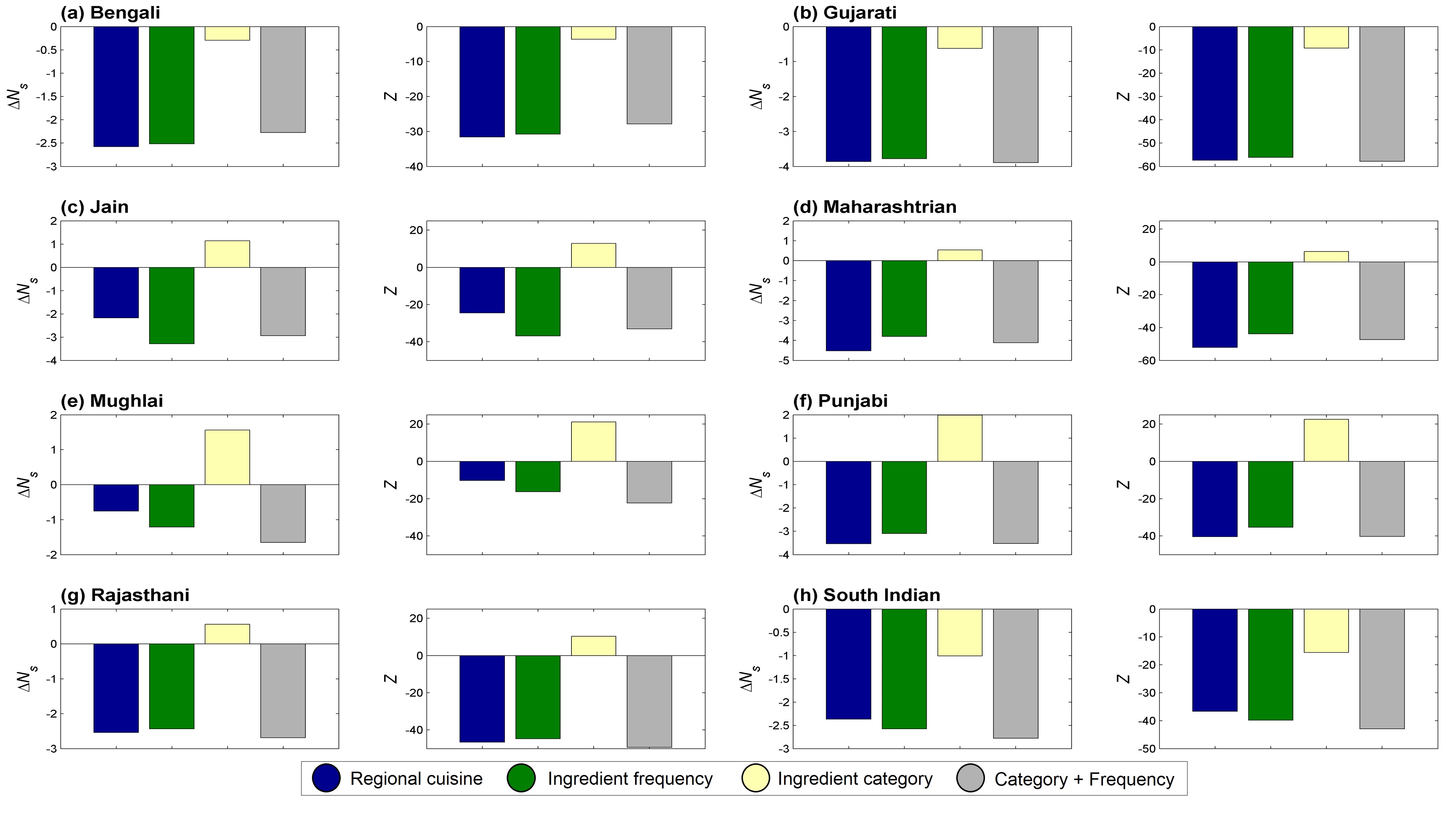}
\caption{{\bf $\Delta N_s$ and its statistical significance.} The variation in $\Delta N_s$ for regional cuisines and corresponding random controls signifying the extent of bias in food pairing. Statistical significance of $\Delta N_s$ is shown in terms of $Z$-score. `Regional cuisine' refers to each of the eight cuisines analyzed; `Ingredient frequency' refers to the frequency controlled random cuisine; `Ingredient category' refers to ingredient category controlling random cuisine; and `Category + Frequency' refers to random control preserving both ingredient frequency and category. Among all regional cuisines, Mughlai cuisine showed least negative food paring ($\Delta N_s = -0.758$) while Maharashtrian cuisine had most negative food pairing ($\Delta N_s = -4.523$).}
\label{Fig_4:delta Ns}
\end{figure*}

\begin{figure*}[ht]
\includegraphics[scale=.43]{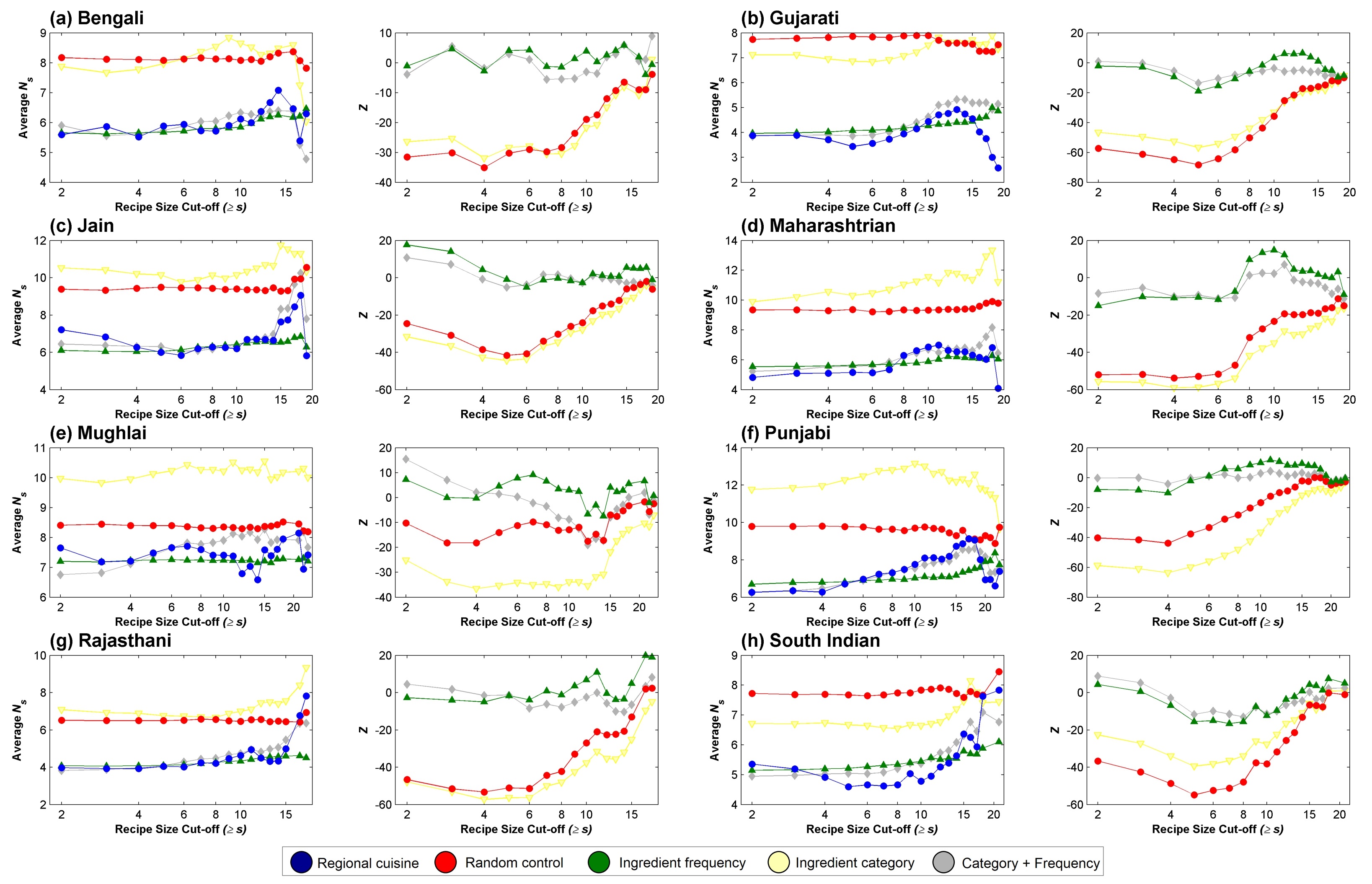}
\caption{{\bf Variation in average $N_s$ and its statistical significance.} 
Change in $\overline{N_s}$ with varying recipe size cut-offs reveals the nature of food pairing across the spectrum of recipe sizes. The $\overline{N_s}$ values for regional cuisines were consistently on the lower side compared to their random counterparts. Category controlled random cuisine displayed average $N_s$ variation close to that of the `Random control'. Frequency controlled as well as `Category + Frequency' controlled random cuisines, on the other hand, displayed average $N_s$ variations close to that of the real-world cuisine.}
\label{Fig_5:average Ns}
\end{figure*}

We found that all regional cuisines are invariantly characterized by average food pairing lesser than expected by chance. This characteristic negative food pairing, however, varied in its extent across cuisines. Mughlai cuisine, for example, displayed the least inclination towards negative pairing ($\Delta N_s = \overline{N}^{Mughlai}_s - \overline{N}^{Rand}_s = -0.758$ and $Z$-score of -10.232). Whereas, Maharashtrian cuisine showed the most negative food pairing ($\Delta N_s = \overline{N}^{Maharashtrian}_s - \overline{N}^{Rand}_s = -4.523$ and $Z$-score of -52.047). Figure~\ref{Fig_4:delta Ns} depicts the generic food pairing pattern observed across regional cuisines of India. We found that the negative food pairing is independent of recipe size as shown in Figure~\ref{Fig_5:average Ns}. This indicates that the bias in food pairing is not an artefact of averaging over recipes of all sizes and is a quintessential feature of all regional cuisines of India. Note that, across cuisines, majority of recipes are in the size-range of around 3 to 12. Hence the significance of food pairing statistics is relevant below the recipe size cut-off of $\sim$12. 

We further investigated for possible factors that could explain negative food pairing pattern observed in regional cuisines. We created randomized controls for each regional cuisine to explore different aspects that may contribute to the bias in food pairing. In the first control, frequency of occurrence of each ingredient was preserved at the cuisine level (`Ingredient frequency'). In the second control, category composition of each recipe was preserved (`Ingredient category'). A third composite control was created by preserving both category composition of each recipe as well as frequency of occurrence of ingredients (`Category + Frequency'). 

Interestingly, ingredient frequency came out to be a critical factor that could explain the observed bias in food pairing as reflected in $\overline{N}_s$ (Figure~\ref{Fig_4:delta Ns}). The pattern of food pairing across different size-range of recipes is also consistent with this observation (Figure~\ref{Fig_5:average Ns}). On the contrary, category composition itself turned out to be irrelevant and led to food pairing that was similar to that of a randomized cuisine. Further, the control implementing a composite model featuring both the above aspects recreated food pairing observed in regional cuisines. Thus frequency of occurrence of ingredients emerged as the most central aspect which is critical for rendering the characteristic food pairing.

\subsection{Food pairing at recipe level}
\begin{figure*}[ht]
\includegraphics[scale=.43]{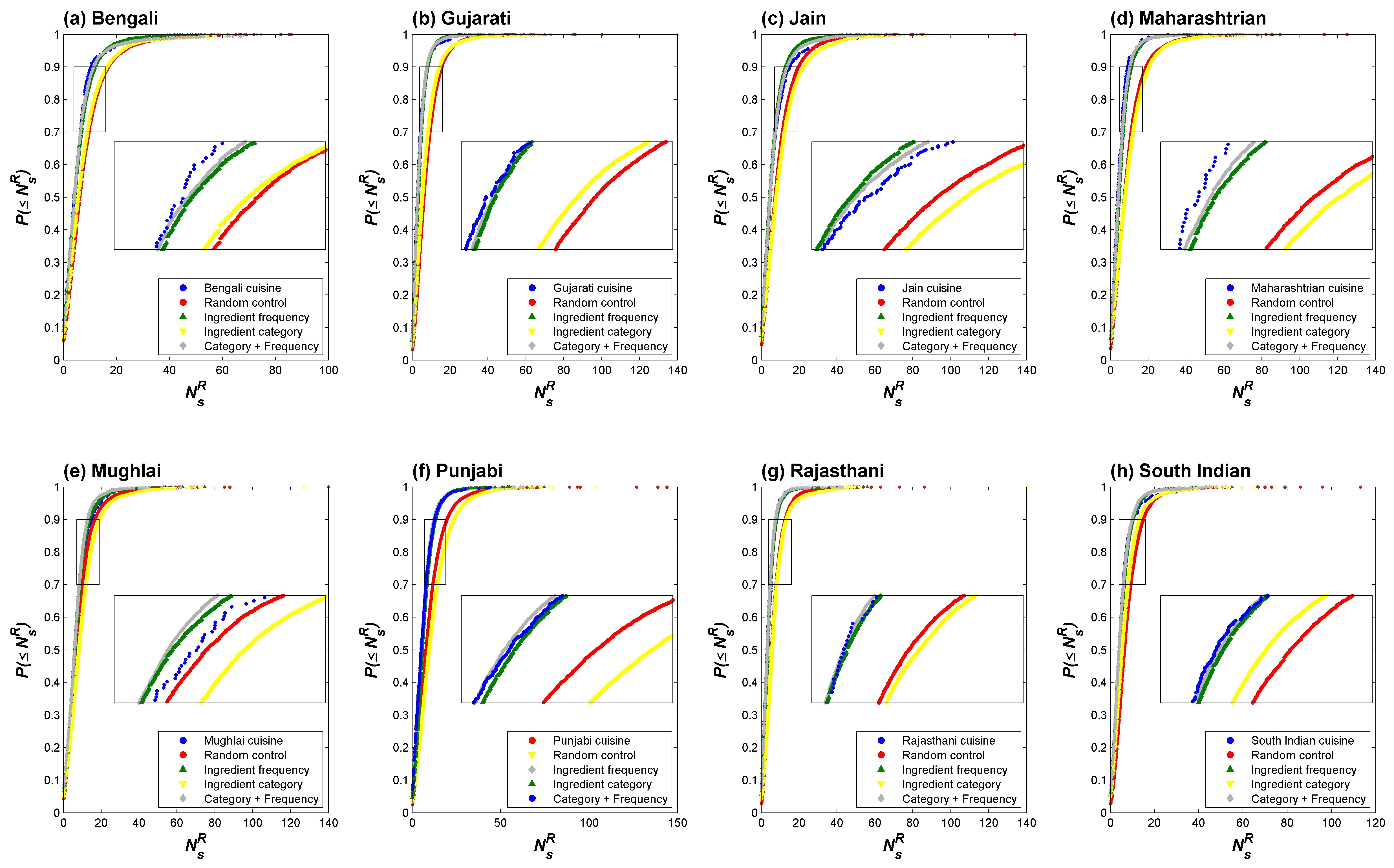}
\caption{{\bf Cumulative probability distribution of $N_s^R$ values for regional cuisines and their random controls.} Cumulative distribution of $N_s^R$ indicates the probability of finding a recipe having food pairing less than or equal to $N_s^R$. The data of regional cuisines as well as those of their controls were fitted with a sigmoid equation indicating that the $P(N_s^R)$ values fall exponentially. The exponent $\alpha$ (Equation~\ref{eq:nsr_cum_distn}) refers to the rate of decay; larger the $\alpha$ more prominent is the negative food pairing in recipes of a cuisine. As evident from~\ref{S2_Table}, $N_s^R$ distribution of the controls based on `Ingredient Frequency' as well as `Category + Frequency' displayed recipe level food pairing similar to real-world cuisines. On the other hand, as also observed at the level of cuisine (Figure~\ref{Fig_4:delta Ns} and Figure~\ref{Fig_5:average Ns}), both the `Random Control' as well as `Ingredient Category' control deviate significantly.}
\label{Fig_6:PNsR vs NsR}
\end{figure*}

Looking into the food pairing at recipe level, we analyzed the nature of distribution of food pairing among recipes ($N_s^R$). Our analysis showed that the negative $\Delta N_s$ observed for cuisines was not an averaging effect. The $N_s^R$ values tend to follow exponential distribution, indicating that number of recipes exponentially decays with increasing $N_s^R$. To address the noise due to small size of cuisines, we computed cumulative distribution ($P(\leq N_s^R)$) as depicted in Figure~\ref{Fig_6:PNsR vs NsR}. The nature of cumulative distribution for an exponential probability distribution function ($P(N_s^R) \propto e^{-\alpha N_s^R}$) would be of the following form: 

\begin{equation}\label{eq:nsr_cum_distn}
P(\leq N_s^R)= a+ \frac{k-a}{1+e^{-\alpha N_s^R}}
\end{equation}

We found that all regional cuisines show a strong bias towards recipes of low $N_s^R$ values as observed in Figure~\ref{Fig_6:PNsR vs NsR}. For each regional cuisine, the bias was accentuated in comparison to corresponding random cuisines as reflected in the exponents shown in~\ref{S2_Table}. Once again Mughlai cuisine emerged as an outlier, as the nature of its $N_s^R$ distribution did not indicate a clear distinction from that of its random control. Consistent with the observation made with $\overline{N}_s$ and $\Delta N_s$ statistics (Figure~\ref{Fig_4:delta Ns} and Figure~\ref{Fig_5:average Ns}), we found that controlling for frequency of occurrence of ingredients reproduces the nature of $N_s^R$ distribution across all regional cuisines (barring the Mughlai cuisine). This further highlights the role of ingredient frequency as a key factor in specifying food pairing at the level of recipes as well.

\subsection{Food pairing at the level of ingredient pairs}
\begin{figure*}[ht]
\includegraphics[scale=.4]{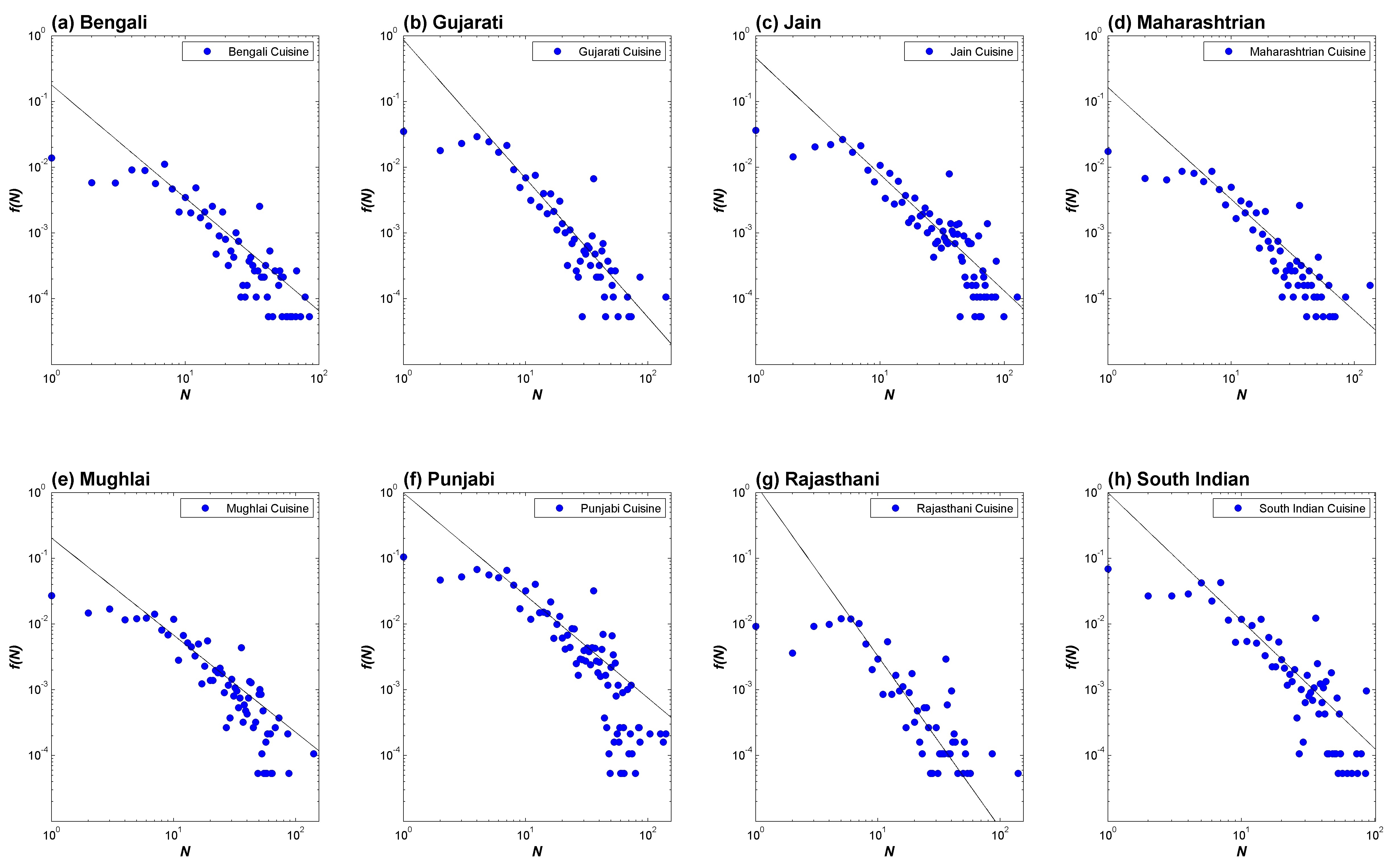}
\caption{{\bf Co-occurrence of ingredients with increasing extent of flavor profile overlap.} Fraction of ingredient pair occurrence ($f(N)$) with a certain extent of flavor profile overlap ($N$) was computed  to assess the nature of food pairing at the level of ingredient pairs. Generically across the cuisines it was observed that, the occurrences of ingredient pairs dropped as a power law with increasing extent of flavor profile sharing. This further ascertained negative food pairing pattern in regional cuisines, beyond the coarse-grained levels of cuisine and recipes.
}
\label{Fig_7:fNs vs Ns}
\end{figure*}

Beyond the level of cuisine and recipes, the bias in food pairing can be studied at the level of ingredient pairs. We computed co-occurrence of ingredients in the cuisine for increasing value of flavor profile overlap ($N$). We found that the fraction of pairs of ingredients with a certain overlap of flavor profiles ($f(N)$) followed a power law distribution $f(N) \propto N^{-\gamma} $ (Figure~\ref{Fig_7:fNs vs Ns}). This indicates that higher the extent of flavor overlap between a pair of ingredients, the lesser is its usage in these cuisines.~\ref{S3_Table} lists the $\gamma$ values for each of the regional cuisines. 

\subsection{Contribution of individual ingredients towards food pairing}
\begin{figure*}[ht]
\includegraphics[scale=.43]{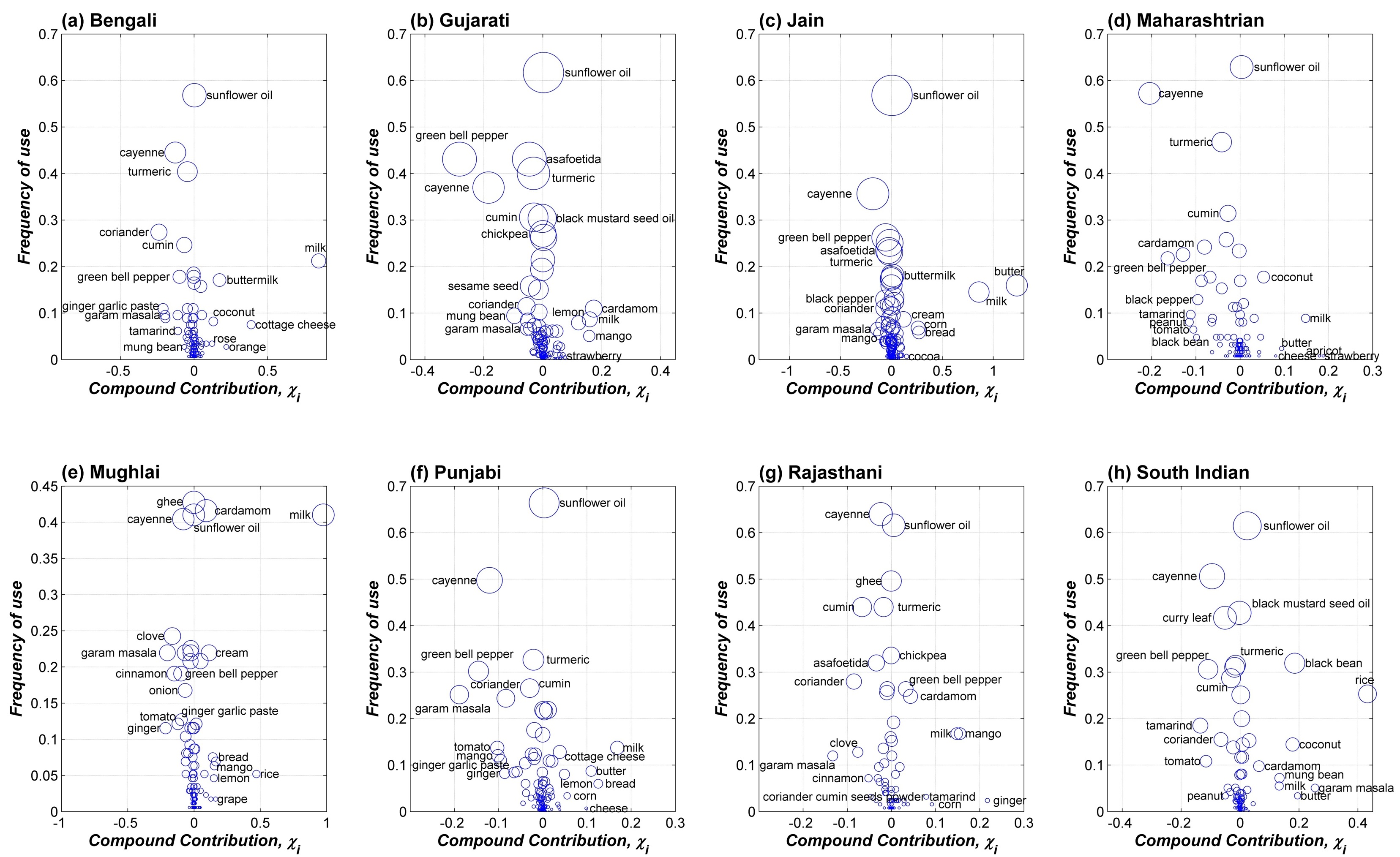}
\caption{{\bf Contribution of ingredients ($\chi_i$) towards flavor pairing.} For all eight regional cuisines we calculated the $\chi_i$ value of ingredients that indicates their contribution to flavor pairing pattern of the cuisine and plotted them against their frequency of appearance. Size of circles are proportional to frequency of ingredients. Across cuisines, prominent negative contributors largely comprised of spices, whereas a few dairy products consistently appeared on the positive side. }
\label{Fig_8:Xi}
\end{figure*}

For each of the regional cuisines we calculated the contribution of ingredients ($\chi_i$) towards the food pairing pattern. For an ingredient whose presence in the cuisine does not lead to any bias, the value of $\chi_i$ is expected to be around zero. With increasing role in biasing food pairing towards positive (negative) side, $\chi_i$ is expected to be proportionately higher (lower). Figure~\ref{Fig_8:Xi} shows the distribution of ingredient contribution ($\chi_i$) and its frequency of occurrence, for each regional cuisine. Ingredients that make significant contribution towards food pairing could be located, in either positive or negative side, away from the neutral vertical axis around $\chi_i=0$. Significantly, spices were consistently present towards the negative side, while milk and certain dairy products were present on the positive side across cuisines. Prominently among the spices, cayenne consistently contributed to the negative food pairing of all  regional cuisines. Certain ingredients appeared to be ambivalent in their contribution to food pairing. While cardamom contributed to the positive food pairing in Gujarati, Mughlai, Rajasthani, and South Indian cuisines, it added to negative food pairing in Maharashtrian cuisine. Green bell pepper tends to contribute to negative food pairing across the cuisines except in the case of Rajasthani cuisine. Details of $\chi_i$ values of prominent ingredients for each regional cuisine are presented in~\ref{S4_Table}. 

\subsection{Role of ingredient categories in food pairing}
\begin{figure*}[ht]
\includegraphics[scale=.43]{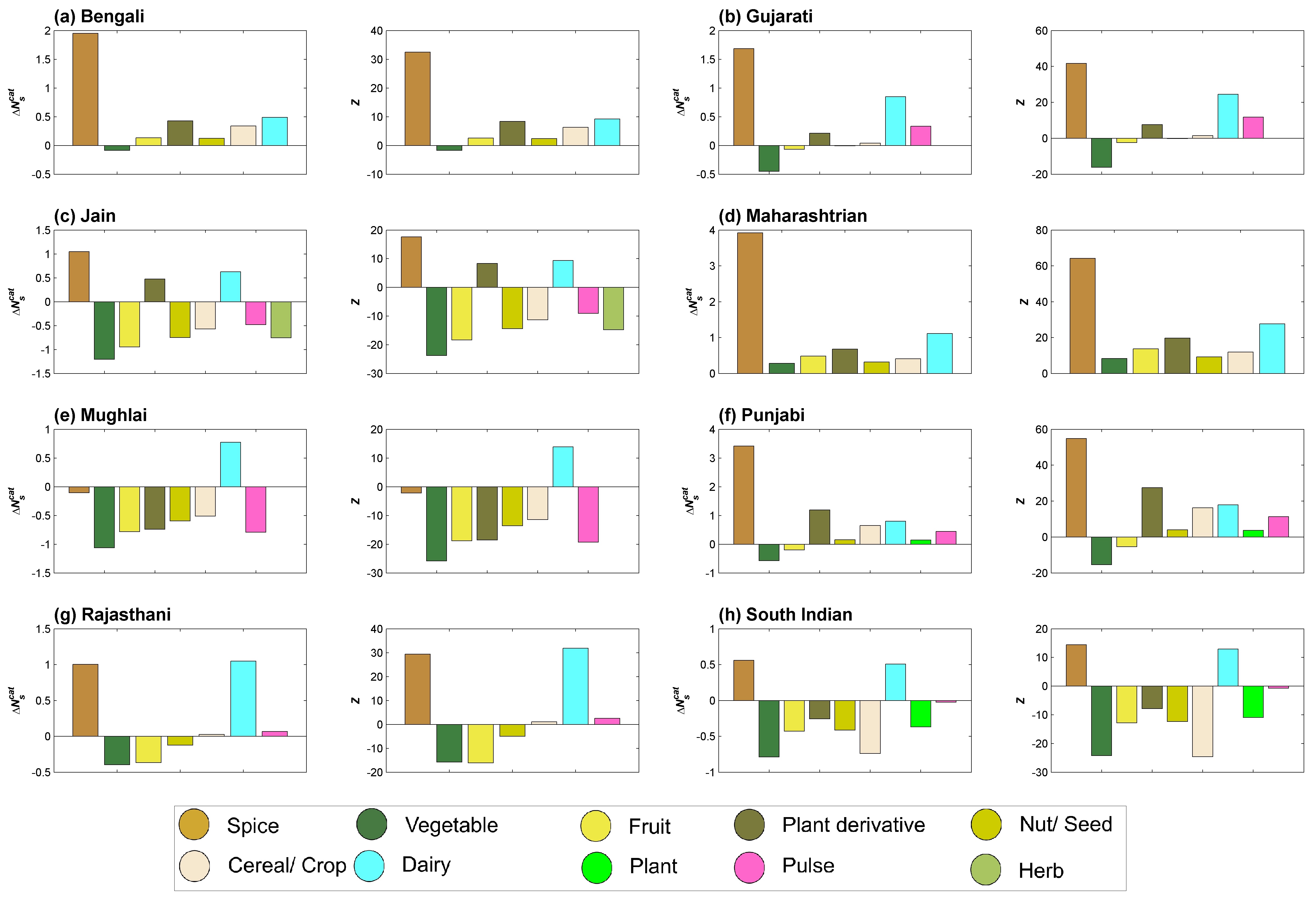}
\caption{{\bf Contribution of individual categories ($\Delta N_s^{cat}$) towards food pairing bias and its statistical significance.} Randomizing ingredients within a certain category provides an insight into their contribution towards bias in food pairing. Spice and dairy category  showed up as prominent categories contributing to the negative food pairing of regional cuisines. 
}
\label{Fig_9:delta Ns cat}
\end{figure*}

As discussed earlier, the random cuisine where only category composition of recipes was conserved, tends to have food pairing similar to that of the `Random control'~(Figure~\ref{Fig_4:delta Ns} and Figure~\ref{Fig_5:average Ns}). This raises the question whether ingredient category has any role in determining food pairing pattern of the cuisine. Towards answering this question, we created random cuisines wherein we randomized ingredients within one category, while preserving the category and frequency distribution for rest of the ingredients. The extent of contribution of an ingredient category towards the observed food pairing in the cuisine is represented by $\Delta N_s^{cat}$. Figure~\ref{Fig_9:delta Ns cat} depicts significance of ingredient categories towards food pairing of each regional cuisine. Interestingly, the pattern of category contributions presents itself as a `culinary fingerprint' of the cuisine.

\begin{figure*}[ht]
\includegraphics[scale=.43]{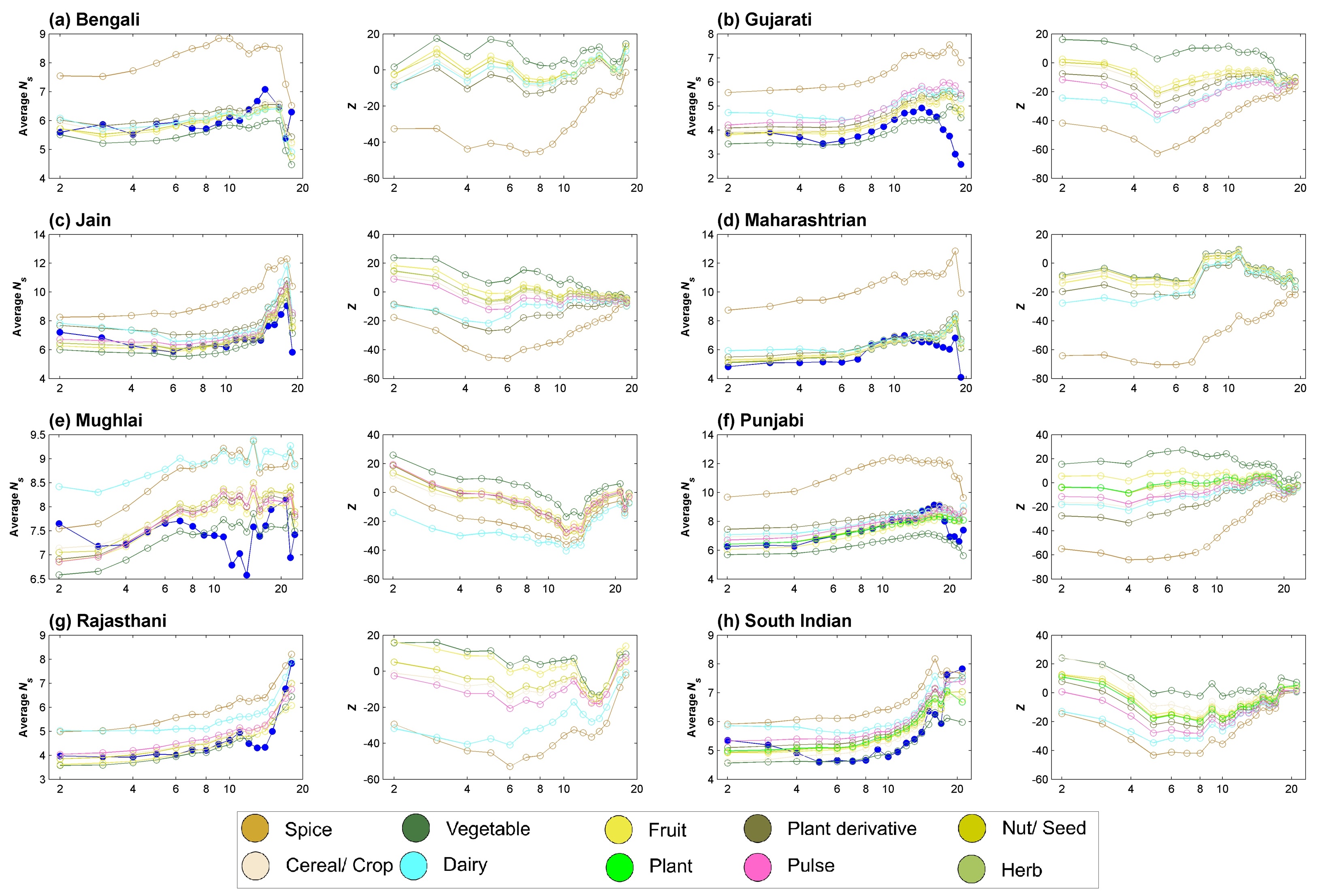}
\caption{{\bf Variation in category contribution and its statistical significance.} Across the spectrum of recipe sizes, we observed broadly consistent trend of contribution of individual categories towards food pairing bias.}
\label{Fig_10:average Ns cat}
\end{figure*}

The `spice' category was the most significant contributor to negative food pairing across cuisines with the exception of Mughlai cuisine. Another category which consistently contributed to negative food pairing was `dairy'. On the other hand, `vegetable' and `fruit' categories tend to bias most cuisines towards positive food pairing. Compared to the above-mentioned categories, `nut/seed', `cereal/crop', `pulse' and `plant derivative' did not show any consistent trend. `Plant' and `herb' categories, sparsely represented in cuisines, tend to tilt the food pairing towards positive side. In Mughlai cuisine all ingredient categories, except `dairy', tend to contribute towards positive food pairing. This could be a reflection of the meagre negative food pairing observed for the cuisine~(Figure~\ref{Fig_4:delta Ns}). Above observations were found to be consistent across the spectrum of recipe sizes~(Figure~\ref{Fig_10:average Ns cat}).

\section{Conclusions}
With the help of data analytical techniques we have shown that food pairing in major Indian regional cuisines follow a consistent trend. We analyzed the reason behind this characteristic pattern and found that spices, individually and as a category, play a crucial role in rendering the negative food pairing to the cuisines. The use of spices as a part of diet dates back to ancient Indus civilization of Indian subcontinent~\cite{Kashyap2010,Lawler2012,Weber2011} . 
They also find mention in Ayurvedic texts such as \emph{Charaka Samhita} and \emph{Bhaavprakash Nighantu}~\cite{AtridevjiGupt1948,Valiathan2010,Pande2002,Tapsell2006}. \emph{Trikatu}, an Ayurvedic formulation  prescribed routinely for a variety of diseases, is a combination of spices viz., long pepper, black pepper and ginger~\cite{Johri1992}. Historically spices have served several purposes such as coloring and flavoring agents, preservatives and additives. They also serve as anti-oxidants, anti-inflammatory, chemopreventive, antimutagenic and detoxifying agents~\cite{Tapsell2006,Krishnaswamy2008a}. One of the strongest hypothesis proposed to explain the use of spices is the antimicrobial hypothesis, which suggests that spices are primarily used due to their activity against food spoilage bacteria~\cite{Billing1998,Sherman1999}. A few of the most antimicrobial spices~\cite{Rakshit2010a} are commonly used in Indian cuisines. Our recent studies have shown the beneficial role of capsaicin, an active component in cayenne which was revealed to be the most prominent ingredient in consistently rendering the negative food pairing in all regional cuisines~\cite{Perumal2014}. The importance of spices in Indian regional cuisines is also highlighted by the fact these cuisines have many derived ingredients (such as garam masala, ginger garlic paste etc.) that are spice combinations. The key role of spices in rendering characteristic food pairing in Indian cuisines and the fact that they are known to be of therapeutic potential, provide a basis for exploring possible causal connection between diet and health as well as prospection of therapeutic molecules from food ingredients.
Flavor pairing has been used as a basic principle in algorithm design for both recipe recommendation and novel recipe generation, thereby enabling computational systems to enter the creative domain of cooking and suggesting recipes~\cite{Varshney2013a,Teng2012}. In such algorithms, candidate recipes are generated based on existing domain knowledge and flavor pairing plays a crucial role while selecting the best among these candidates~\cite{Teng2012}.

\section{Materials and Methods}
\subsection{Data collection and curation}
The data of regional cuisines were obtained from one of the leading cookery websites of Indian cuisine, tarladalal.com (December 2014). Among various online resources available for Indian cuisine, \emph{TarlaDalal}~\cite{Dalal2014} (http://www.tarladalal.com) was found to be the best in terms of authentic recipes, cuisine annotations and coverage across major regional cuisines. The website had 3330 recipes from 8 Indian cuisines. Among others online sources: Sanjeev Kapoor (http://www.sanjeevkapoor.com) had 3399 recipes from 23 Indian cuisines; NDTV Cooks (http://cooks.ndtv.com) had 667 Indian recipes across 15 cuisines; Manjula’s Kitchen (http://www.manjulaskitchen.com) was restricted to 730 Indian vegetarian recipes across 19 food categories; Recipes Indian (http://www.recipesindian.com) had 891 recipes from around 16 food categories; All Recipes (http://www.allrecipes.com) had only 449 recipes from 6 food categories. In comparison to these sources, Tarladalal.com was identified as a best recipe source of Indian cuisine.

The data of 3330 recipes and 588 ingredients were curated for redundancy in names and to drop recipes with only one ingredient. These ingredients belonged to 17 categories. Ingredients of `snack' and `additive' categories, for which no flavor compounds could be determined, were removed. The ingredients were further aliased to 339 source ingredients out of which we could determine flavor profiles for 194 of them. Aliasing involves mapping ingredients to their source ingredient. For example `chopped potato' and `mashed potato' were aliased to `potato'. The final data comprised of 2543 recipes and 194 ingredients belonging to 15 categories. The statistics of regional cuisines, their recipes and ingredient counts is provided in Table~\ref{table1}.

The data of flavor compounds were obtained from Ahn et.\ al.\ \cite{Ahn2011}, Fenaroli's Handbook of Flavor Compounds~\cite{Burdock2010} and extensive literature search. All the flavor profiles were cross checked with those in 6th edition (latest) of Fenaroli’s Handbook of Flavor Compounds~\cite{Burdock2010} for consistency of names. Chemical Abstract Service numbers were used as unique identifiers of flavor molecules.

\subsection{Flavor sharing}
Flavor sharing was computed for each pair of ingredients that co-occur in recipes in terms of number of shared compounds $N = |F_i \cap F_j| $. Further, the average number of shared compounds in a recipe $N_s^R$ having $s$ ingredients was calculated (Equation~\ref{eq:nsr}).

\begin{equation}\label{eq:nsr}
N_s^R = \frac{2}{s(s-1)} \sum_{i,j\in R, i \neq j} |F_i \cap F_j|
\end{equation}
where $F_i$ represents the flavor profile of ingredient $i$ and $R$ represents a recipe.

For a cuisine with $N_R$ recipes, we then calculated the average flavor sharing of the cuisine $\overline{N}_s^{cuisine} (= \frac{\Sigma_R N_s^R}{N_R})$. Figure~\ref{Fig_3:Ns calculation} illustrates this procedure graphically. We compared average $N_s$ of the cuisine with that of corresponding randomized cuisine (Figure~\ref{Fig_4:delta Ns}) by calculating $\Delta N_s$ ($=\overline{N}_s^{cuisine} - \overline{N}_s^{Rand}$), where $cuisine$ and $Rand$ indicate the regional cuisine and corresponding `random cuisine' respectively.

A total of four random controls were created viz. `Random control', `Ingredient frequency', `Ingredient category' and `Category + Frequency'. While in all random cuisines recipe size distribution of the original cuisine was preserved, `Random control' implemented uniform selection of ingredients (1 set of 10,000 recipes for each regional cuisine); `Ingredient frequency' control was created while maintaining the ingredient usage frequency distribution (1 set of 10,000 recipes for each regional cuisine); `Ingredient category' control was created by randomizing ingredient usage in recipes with ingredients belonging to same categories, thus maintaining the category composition of recipes (8 sets of recipes for a total of $>10,000$ recipes for each regional cuisine); and `Category + Frequency' control preserved both the ingredient categories in recipes as well as frequency of overall ingredient usage within the cuisine (8 sets of recipes for a total of $>10,000$ recipes for each regional cuisine).

The statistical significance of $\overline{N}_s$ and $\Delta N_s$ was measured with corresponding Z-scores given by 
\begin{equation}
\label{eq:z-score}
Z = \sqrt{N_{Rand}} \frac{(\overline{N}_s^{cuisine} - \overline{N}_s^{Rand})}{\sigma_{Rand}},
\end{equation}
where $N_{Rand}$ and $\sigma_{Rand}$ represent the number of recipes in randomized cuisine and standard deviation of $N_s^R$ values for randomized cuisine respectively.

\subsection{Ingredient contribution}
For every regional cuisine, the contribution ($\chi_i$) of each ingredient $i$ was calculated~\cite{Ahn2011} using Equation~\ref{eq:Xi}.

\begin{eqnarray}
\label{eq:Xi}
\chi_i = \left( \frac{1}{N_R} \sum_{i \in R} \frac{2}{n(n-1)} \sum_{j \neq i (j,i \in R)} |F_i \cap F_j|\right) -\nonumber\\
\left(\frac{2f_i}{N_R \langle n \rangle} \frac{\Sigma_{j \in c} f_j |F_i \cap F_j|}{\Sigma_{j \in c} f_j}\right),
\end{eqnarray}

Here, $f_i$ is the frequency of occurrence of ingredient $i$.

$\chi_i$ values reflect the extent of an ingredient's contribution towards positive or negative food pairing of the cuisine.

\subsection{Uniqueness of ingredient category}
Despite significant flavor sharing within each category of ingredients, the uniqueness of each category, by virtue of combination of its ingredients with other ingredients, was enumerated by intra-category randomization. The average food pairing of such cuisine, randomized for a category, was compared with that of the original cuisine. Such category-randomized cuisines were created only for major categories (having 5 or more ingredients) within each regional cuisine. The deviation in $\overline{N}_s^{cat}$, that reflects the relevance of unique placements of ingredients of $cat$, was calculated using Equation~\ref{eq:delta Ns cat}.
\begin{equation}
\label{eq:delta Ns cat}
\Delta N_s^{cat} = \overline{N}_s^{cat} - \overline{N}_s^{cuisine},  \forall s \geq 2
\end{equation}
Here, $cat$ stands for an ingredient category and $s$ represents recipe size. The statistical significance was again calculated using Z-score.

\section{Supporting Information}
\subsection{S1 Table}
\label{S1_Table}
{\bf Distribution of ingredients across categories.} 
Number of ingredients in each category for all regional cuisines.

\subsection{S2 Table}
\label{S2_Table}
{\bf Exponents ($\alpha$) of Sigmoid fits for $P(N_s^R)$ vs $N_s^R$ distribution.} 
Exponents ($\alpha$) for regional cuisines and their random controls.

\subsection{S3 Table}
\label{S3_Table}
{\bf Power law exponents ($\gamma$) for $f(N)$ vs $N$ distribution.} 
Power law exponents ($\gamma$) of all regional cuisines.

\subsection{S4 Table}
\label{S4_Table}
{\bf Ingredients contributing significantly to food pairing.} 
Details of top 10 ingredients contributing to positive and negative food pairing in each of the regional cuisines.

\begin{acknowledgments}
G.B. acknowledges the seed grant support from Indian Institute of Technology Jodhpur (IITJ/SEED/2014/0003). A.J. and R.N.K. thank the Ministry of Human Resource Development, Government of India as well as Indian Institute of Technology Jodhpur for scholarship and Junior Research Fellowship, respectively.

\end{acknowledgments}



\pagebreak
\widetext
\begin{center}
\textbf{\large Supporting Information: Analysis of food pairing in regional cuisines of India}
\end{center}
\setcounter{section}{0}
\setcounter{equation}{0}
\setcounter{figure}{0}
\setcounter{table}{0}
\setcounter{page}{1}
\makeatletter
\renewcommand{\theequation}{S\arabic{equation}}
\renewcommand{\thefigure}{S\arabic{figure}}

\section{Supporting Tables}

\subsection{S1 Table}
\begin{table}[!h]
\caption{
{\bf Distribution of ingredients across categories.} Number of ingredients in each category for all regional cuisines.}
\begin{tabular}{|l|r|r|r|r|r|r|r|r|}
\hline
\multicolumn{1}{|l|}{\bf Ingredient Category} & \multicolumn{1}{l|}{\bf Bengali} & \multicolumn{1}{l|}{\bf Gujarati} & \multicolumn{1}{l|}{\bf Jain} & \multicolumn{1}{l|}{\bf Maharashtrian} & \multicolumn{1}{l|}{\bf Mughlai} & \multicolumn{1}{l|}{\bf Punjabi} & \multicolumn{1}{l|}{\bf Rajasthani} & \multicolumn{1}{l|}{\bf South Indian} \\ \hline
spice & 25 & 23 & 26 & 25 & 24 & 33 & 21 & 25 \\ \hline
vegetable & 14 & 23 & 29 & 14 & 15 & 29 & 16 & 23 \\ \hline
fruit & 13 & 19 & 25 & 9 & 16 &	22 & 5 & 14  \\ \hline
plant derivative & 8 & 7 & 11 & 7 &	8 &	13 & 4 & 6  \\ \hline
nut/seed & 12 & 12 & 12 & 11 & 11 & 13 & 8 & 10  \\ \hline
cereal/crop & 6 & 10 & 11 & 6 & 9 & 12 & 7 & 9 \\ \hline
dairy & 7 & 6 & 8 &	6 &	7 &	10 & 5 & 7  \\ \hline
plant &	2 &	3 &	3 &	3 &	4 &	5 &	4 &	5  \\ \hline
pulse & 4 &	6 &	5 &	4 &	5 &	6 &	5 &	6  \\ \hline
herb &	2 & 2 &	5 &	3 &	3 &	4 &	2 &	3  \\ \hline
meat &	3 &	0 &	0 &	2 &	0 &	1 &	0 &	0  \\ \hline
beverage & 1 & 0 & 1 & 1 & 0 & 1 & 0 & 0  \\ \hline
fish/seafood & 2 & 0 & 0 & 0 & 0 & 0 & 0 & 2  \\ \hline
animal product & 2 & 0 & 1 & 1 & 2 & 2 & 0 & 2  \\ \hline
flower & 1 & 1 & 1 & 1 & 1 & 1 & 1 & 1  \\ \hline
additive & 0 & 0 & 0 & 0 & 0 & 0 & 0 & 1  \\ \hline
\end{tabular}
\label{s1_table_sup}
\end{table}

\subsection{S2 Table}
\begin{table}[!h]
\caption{
{\bf Exponents $(\alpha)$ of sigmoid fits for $P(N_s^R)$ vs $N_s^R$ distribution.} Exponents ($\alpha$) for regional cuisines and their random controls.}
\begin{tabular}{|l|r|r|r|r|r|}
\hline
\multirow{2}{*}{\bf Cuisine} & \multicolumn{5}{c|}{\bf $\alpha$ Values}  \\
\cline{2-6}
& {\bf Original} & {\bf Random control}	& {\bf Ingredient frequency} & {\bf Ingredient category} &	{\bf Category + Frequency} \\  \hline
Bengali & 0.255525 & 0.181436 & 0.255149 & 0.190506 & 0.26209 \\ \hline
Gujarati & 0.405862 & 0.187475 & 0.365109 &	0.207978 & 0.37633 \\ \hline
Jain &	0.226656 &	0.155991 &	0.235283 &	0.138507 &	0.228731 \\ \hline
Maharashtrian &	0.282265 &	0.158809 &	0.259422 &	0.141178 &	0.269226 \\ \hline
Mughlai &	0.184891 &	0.173672 &	0.202563 &	0.143178 &	0.194965 \\ \hline
Punjabi &	0.207118 &	0.150068 &	0.207771 &	0.120212 &	0.215736 \\ \hline
Rajasthani &	0.315478 &	0.223507 &	0.35912 &	0.209513 &	0.351726 \\ \hline
South Indian &	0.300892 &	0.189509 &	0.280907 &	0.213137 &	0.290387 \\ \hline
\end{tabular}
\label{s2_table_sup}
\end{table}

\pagebreak

\subsection{S3 Table}
\begin{table}[!h]
\caption{
{\bf Power law exponent ($\gamma$) for $f(N)$ v/s $N$ distribution.} Power law exponent ($\gamma$) for all regional cuisines.}
\begin{tabular}{|l|r|}
\hline
\multicolumn{1}{|l|}{\bf Cuisine} & \multicolumn{1}{|l|}{\bf $\gamma$ Value} \\ \hline
Bengali & 1.71906 \\ \hline
Gujarati & 2.11136 \\ \hline
Jain & 1.77156 \\ \hline
Maharashtrian & 1.6974 \\ \hline
Mughlai & 1.47354 \\ \hline
Punjabi & 1.55844 \\ \hline
Rajasthani & 2.62489 \\ \hline
South Indian & 1.948 \\ \hline
\end{tabular}
\label{s3_table_sup}
\end{table}

\subsection{S4 Table}
\begin{longtable}[!ht]{| m{14em} | R{5em} | R{7em} | m{14em} | R{5em} | R{7em} |}

\caption{
{\bf Ingredients contributing significantly to food pairing.} Details of top 10 ingredients contributing to positive and negative food pairing in each of the regional cuisines.} \label{s4_table_sup}\\
\hline
\multicolumn{6}{|c|}{\bf Bengali} \\ \hline
{\bf Ingredients contributing to negative food pairing} & \multicolumn{1}{r|}{\bf $\chi$ value} & {\bf Frequency of occurrence} & {\bf Ingredients contributing to positive food pairing} & {\bf $\chi$ value} & {\bf Frequency of occurrence} \\ \hline

coriander &	-0.24319 &	40 & milk &	0.84165 & 31 \\ \hline
ginger garlic paste & -0.21437 & 16 & cottage cheese & 0.38636 &	11 \\ \hline
garam masala &	-0.20126 &	14 & orange & 0.21789 &	4 \\ \hline
mango & -0.19701 & 13 & buttermilk & 0.17259 & 25 \\ \hline
cayenne & -0.13469 & 65 & coconut & 0.13006 & 12 \\ \hline
tomato & -0.11413 & 14 & rose & 0.12178 & 5 \\ \hline
tamarind & -0.11053 & 9 & cocoa & 0.08218 & 5 \\ \hline
green bell pepper & -0.10233 & 26 & strawberry & 0.05512 & 2 \\ \hline
cumin & -0.06875 & 36 & cream & 0.05368 & 5 \\ \hline
mung bean & -0.06702 & 4 & saffron & 0.05329 & 14 \\ \hline

\multicolumn{6}{|c|}{\bf Gujarati} \\ \hline
{\bf Ingredients contributing to negative food pairing} & {\bf $\chi$ value} & {\bf Frequency of occurrence} & {\bf Ingredients contributing to positive food pairing} & {\bf $\chi$ value} & {\bf Frequency of occurrence} \\ \hline

green bell pepper & -0.29066 & 169 & cardamom & 0.17035 & 43 \\ \hline
cayenne & -0.19164 & 145 & milk & 0.158002 & 34 \\ \hline
mung bean & -0.09783 & 37 & mango & 0.15628 & 20 \\ \hline
coriander & -0.05721 & 45 & lemon & 0.11942 & 31 \\ \hline 
garam masala & -0.05695 & 26 & strawberry & 0.07485 & 2 \\ \hline 
black pepper & -0.05281 & 33 & chaat masala & 0.06775 & 4 \\ \hline 
asafoetida & -0.04863 & 169 & apple & 0.06058 & 2 \\ \hline 
coriander cumin seeds powder & -0.04469 & 26 & mint & 0.05999 & 11 \\ \hline 
sesame seed & -0.04148 & 62 & apricot & 0.05948 & 1 \\ \hline 
Turmeric & -0.03435 & 157 & cottage cheese & 0.05743 & 4 \\ \hline

\multicolumn{6}{|c|}{\bf Jain} \\ \hline
{\bf Ingredients contributing to negative food pairing} & {\bf $\chi$ value} & {\bf Frequency of occurrence} & {\bf Ingredients contributing to positive food pairing} & {\bf $\chi$ value} & {\bf Frequency of occurrence} \\ \hline

cayenne & -0.18622 & 152 & butter & 1.22722 & 68 \\ \hline 
garam masala & -0.14199 & 28 & milk & 0.85545 & 62 \\ \hline 
mango & -0.11421 & 24 & bread & 0.26881 & 25 \\ \hline 
black bean & -0.08291 & 33 & corn & 0.26018 & 29 \\ \hline 
coriander & -0.06855 & 47 & cocoa & 0.14714 & 3 \\ \hline 
tamarind & -0.06793 & 17 & cream & 0.11764 & 37 \\ \hline 
black pepper & -0.06234 & 55 & peanut butter & 0.09925 & 4 \\ \hline 
green bell pepper & -0.06095 & 112 & grape & 0.09078 & 4 \\ \hline 
ginger & -0.06059 & 17 & cheese & 0.08762 & 11 \\ \hline 
chaat masala & -0.05613 & 14 & strawberry & 0.08254 & 4 \\ \hline

\multicolumn{6}{|c|}{\bf Maharashtrian} \\ \hline
{\bf Ingredients contributing to negative food pairing} & {\bf $\chi$ value} & {\bf Frequency of occurrence} & {\bf Ingredients contributing to positive food pairing} & {\bf $\chi$ value} & {\bf Frequency of occurrence} \\ \hline

cayenne & -0.20961 & 71 & strawberry & 0.18767 & 1 \\ \hline 
green bell pepper & -0.16631 & 27 & apricot & 0.17937 & 1 \\ \hline 
cardamom & -0.13171 & 28 & milk & 0.14751 & 11 \\ \hline 
peanut & -0.11527 & 10 & butter & 0.09349 & 3 \\ \hline 
tamarind & -0.11284 & 12 & cheese & 0.08038 & 1 \\ \hline 
tomato & -0.10687 & 8 & coconut & 0.05239 & 22 \\ \hline 
black bean & -0.09923 & 6 & sesame seed & 0.04636 & 6 \\ \hline 
black pepper & -0.09723 & 16 & cream & 0.04274 & 2 \\ \hline 
cinnamon & -0.08889 & 21 & cocoa & 0.04255 & 1 \\ \hline 
coriander & -0.08271 & 30 & rice & 0.03092 & 11 \\ \hline

\multicolumn{6}{|c|}{\bf Mughlai} \\ \hline
{\bf Ingredients contributing to negative food pairing} & {\bf $\chi$ value} & {\bf Frequency of occurrence} & {\bf Ingredients contributing to positive food pairing} & {\bf $\chi$ value} & {\bf Frequency of occurrence} \\ \hline

ginger & -0.22264 & 20 & milk & 0.95554 & 71 \\ \hline 
garam masala & -0.22203 & 38 & rice & 0.46744 & 9 \\ \hline 
clove & -0.1727 & 42 & bread & 0.16189 & 12 \\ \hline 
cinnamon & -0.15605 & 33 & grape & 0.16132 & 3 \\ \hline 
tomato & -0.13042 & 21 & mango & 0.14838 & 11 \\ \hline 
ginger garlic paste & -0.10488 & 22 & lemon & 0.14672 & 8 \\ \hline 
green bell pepper & -0.10483 & 33 & chaat masala & 0.13532 & 13 \\ \hline 
cayenne & -0.09472 & 70 & honey & 0.12645 & 3 \\ \hline 
coriander & -0.07582 & 38 & cream & 0.10899 & 38 \\ \hline 
onion & -0.0696 & 29 & soybean & 0.08769 & 4 \\ \hline

\multicolumn{6}{|c|}{\bf Punjabi} \\ \hline
{\bf Ingredients contributing to negative food pairing} & {\bf $\chi$ value} & {\bf Frequency of occurrence} & {\bf Ingredients contributing to positive food pairing} & {\bf $\chi$ value} & {\bf Frequency of occurrence} \\ \hline

garam masala & -0.18891 & 251 & milk & 0.16846 & 137 \\ \hline 
green bell pepper & -0.14559 & 301 & bread & 0.12552 & 60 \\ \hline 
cayenne & -0.1208 & 496 & butter & 0.10934 & 87 \\ \hline 
tomato & -0.10311 & 137 & cheese & 0.09834 & 7 \\ \hline 
mango & -0.10147 & 120 & corn & 0.05484 & 34 \\ \hline 
ginger garlic paste & -0.09551 & 110 & lemon & 0.0488 & 80 \\ \hline 
ginger & -0.08621 & 82 & cottage cheese & 0.03844 & 128 \\ \hline 
coriander & -0.08364 & 243 & grape & 0.03832 & 4 \\ \hline 
cinnamon & -0.06514 & 84 & honey & 0.03591 & 11 \\ \hline 
clove & -0.05827 & 86 & olive & 0.03388 & 16 \\ \hline

\multicolumn{6}{|c|}{\bf Rajasthani} \\ \hline
{\bf Ingredients contributing to negative food pairing} & {\bf $\chi$ value} & {\bf Frequency of occurrence} & {\bf Ingredients contributing to positive food pairing} & {\bf $\chi$ value} & {\bf Frequency of occurrence} \\ \hline

garam masala & -0.13817 & 15 & ginger & 0.21659 & 3 \\ \hline 
coriander & -0.0901 & 35 & mango & 0.15163 & 21 \\ \hline 
clove & -0.07852 & 16 & milk & 0.14564 & 21 \\ \hline 
cumin & -0.07138 & 55 & corn & 0.09148 & 2 \\ \hline 
cinnamon & -0.05325 & 9 & tamarind & 0.07795 & 4 \\ \hline 
coriander cumin seeds powder & -0.04782 & 4 & cardamom & 0.03735 & 31 \\ \hline 
asafoetida & -0.03663 & 40 & butter & 0.03672 & 2 \\ \hline 
cayenne & -0.03646 & 80 & lemon & 0.02806 & 3 \\ \hline 
potato & -0.03488 & 3 & bread & 0.02767 & 2 \\ \hline 
black pepper & -0.03262 & 9 & green bell pepper & 0.02621 & 33 \\ \hline

\multicolumn{6}{|c|}{\bf South Indian} \\ \hline
{\bf Ingredients contributing to negative food pairing} & {\bf $\chi$ value} & {\bf Frequency of occurrence} & {\bf Ingredients contributing to positive food pairing} & {\bf $\chi$ value} & {\bf Frequency of occurrence} \\ \hline

tamarind & -0.13638 & 87 & rice & 0.43068 & 119 \\ \hline 
tomato & -0.11714 & 51 & garam masala & 0.25363 & 24 \\ \hline 
green bell pepper & -0.11087 & 144 & butter & 0.19469 & 16 \\ \hline 
cayenne & -0.09829 & 238 & black bean & 0.1833 & 150 \\ \hline 
coriander & -0.06636 & 73 & coconut & 0.17749 & 68 \\ \hline 
curry leaf & -0.05268 & 196 & mung bean & 0.13281 & 34 \\ \hline 
peanut & -0.05027 & 16 & milk & 0.13233 & 26 \\ \hline 
ginger & -0.04228 & 24 & cardamom & 0.06319 & 46 \\ \hline 
lemon & -0.03363 & 20 & soybean & 0.04396 & 8 \\ \hline 
cumin & -0.03177 & 135 & onion & 0.0302 & 72 \\ \hline


\end{longtable}

\end{document}